\documentclass[12pt]{article}

\usepackage{amsmath,amsfonts,latexsym,amssymb,amscd}
\usepackage{pslatex}
\usepackage[latin1]{inputenc}
\usepackage[T1]{fontenc}
\usepackage{pspicture}
\usepackage{verbatim,amsthm,curves,graphics}
\usepackage{mathrsfs}

\usepackage{graphicx}

\newcommand{\mr}{{\mathbb R}}

\begin{document}


\title{ Uniqueness of the static spacetimes with a photon sphere in Einstein-scalar field theory }

\author{
Stoytcho Yazadjiev$^{}$\thanks{\tt yazad@phys.uni-sofia.bg}
\\ \\
{\it $ ^1$Department of Theoretical Physics, Faculty of Physics, Sofia
University} \\
{\it 5 J. Bourchier Blvd., Sofia 1164, Bulgaria} \\
{\it $ ^2$Theoretical Astrophysics, Eberhard-Karls University of T\"ubingen,}\\ {\it T\"ubingen 72076, Germany  }  }

\date{}

\maketitle

\begin{abstract}
In the present paper we prove a uniqueness theorem for the static and asymptotically flat solutions to the Einstein-scalar field equations which possess a photon sphere.
We show that   such solutions are uniquely specified by their mass $M$ and scalar charge $q$ and that they are isometric  to
the Janis-Newman-Winicour solution with the same mass and scalar charge subject to the inequality $\frac{q^2}{M^2}<3$.
\end{abstract}


\sloppy

\section{Introduction}

General relativity  and the generalized gravitational theories predict the existence of spacetime regions where  light can be confined
in closed orbits \cite{Iyer1985}-\cite{Baldiotti2014}. These regions called photon spheres play a very important role in gravitational lensing which is a standard and powerful tool in
modern observational astronomy and astrophysics \cite{Virbhadra2000},\cite{Virbhadra2002}. From the point of view of  gravitational lensing the photon sphere can be considered
as a timelike hypersurface on which the bending angle of a light ray is unboundedly large \cite{Virbhadra2000},\cite{Virbhadra2002}. A more precise definition will be given below. In the presence of a photon sphere
the gravitational lensing effect gives rise to the so-called relativistic images which are characteristic for  relativistic gravity. It is commonly expected that  ultracompact
objects such as black holes, neutron and boson stars, wormholes and naked singularities are surrounded by a photon sphere \cite{Iyer1985}-\cite{Baldiotti2014}. Apart from its significance for  gravitational lensing the photon sphere is intimately related to the quasinormal modes of  ultracompact objects and is crucially relevant for their stability \cite{Cardoso2014}, \cite{Dafermos2008}-\cite{Decanini2010}.

The photon sphere possesses some very specific characteristics  which make it a very special  and unique object. For example, in static spacetimes, as the explicit exact or numerical solutions show, the lapse function is constant on the photon sphere. Moreover, the photon sphere is a totally  umbilic hypersurface with  constant mean   curvature and constant surface gravity \cite{Claudel2001}, \cite{Foertsch2003}, \cite{Cederbaum2014}. With these properties the photon sphere resembles very much the black hole horizon. It is well known that the presence of an event horizon allows us to classify the asymptotically flat spacetimes only in terms of their conserved asymptotic charges, such  as the mass and the electric charge in the static case
\cite{Heusler1996}. In analogy with the black hole case, the natural question that arises is: {\it Does the presence of a photon sphere uniquely specify the spacetimes with given asymptotic charges?} In other words, we ask whether the spacetimes possessing a photon sphere can be classified only in terms of  asymptotic charges (or other  global physical quantities).  The problem set in this way seems to be much more difficult
that the classification of black hole spacetimes since the spectrum of spacetimes possessing a photon sphere is much  richer than the spectrum of  black hole spacetimes.
Nevertheless, in some cases the desired classification can be achieved. Recently Cederbaum \cite{Cederbaum2014} proved that the static, asymptotically flat solutions to the  vacuum Einstein equations with mass $M$ possessing a photon sphere are isometric to the Schwarzschild solution with the same mass.  In the present paper we consider the more general case of static Einstein-scalar field equations. We prove that the static and the asymptotically flat solutions to the  Einstein-scalar field equations possessing a photon sphere are uniquely specified
by their mass $M$ and scalar charge $q$. More precisely we prove that the static and the asymptotically flat solutions to the  Einstein-scalar field equations  with mass $M$ and scalar charge $q$ possessing a photon sphere are isometric to the Janis-Newman-Winicour solution \cite{Janis1968}  with the same mass and scalar charge as well as with a certain restriction on the ratio $q/M$.

\section{General definitions and equations}

In the present paper we consider Einstein-scalar field theory described by the action

\begin{eqnarray}
{\cal S} = \frac{1}{16\pi}\int d^4x \sqrt{g^{(4)}} \left(R^{(4)} - 2g^{(4)\mu\nu}\nabla^{(4)}_{\mu}\varphi\nabla^{(4)}_{\nu}\varphi \right),
\end{eqnarray}
where $\varphi$ is the scalar field, $\nabla^{(4)}_{\mu}$ and $R^{(4)}$ are the Levi-Civita connection and the Ricci scalar curvature with respect to the spacetime metric
$g_{\mu\nu}$. This action gives rise to the following field equations on the spacetime manifold $M^{(4)}$

\begin{eqnarray}\label{ESFE}
&&Ric^{(4)}_{\,\mu\nu}=2\nabla^{(4)}_{\mu}\varphi\nabla^{(4)}_{\nu}\varphi, \\
&&\nabla^{(4)}_{\mu}\nabla^{(4)\mu}\varphi=0 \nonumber,
\end{eqnarray}
with $Ric^{(4)}_{\,\mu\nu}$ being the Ricci tensor. It is important to note the following. Although the Einstein-scalar field equations can be considered in their own right
they  are in fact  the "vacuum" field equations of scalar-tensor theories presented in the so-called Einstein frame. In this way all results  for (\ref{ESFE}) obtained
in the present paper can be easily transformed as results for an arbitrary class of scalar-tensor theories.

In the present work we focus on static and asymptotically flat spacetimes. A spacetime is called static if there exists a smooth Riemannian manifold $(M^{(3)}, g^{(3)})$ and
a smooth lapse function $N: M^{(3)} \to {\mr}^{+}$ such that

\begin{eqnarray}
M^{(4)}= {\mr} \times M^{(3)}, \,\,\, g= - N^2 dt^2 + g^{(3)}.
\end{eqnarray}
In addition to the metric staticity we  have to defined the scalar field staticity. The scalar field is called static if ${\cal L}_{\xi}\varphi= 0$ where ${\cal L}_{\xi}$ is
the Lie derivative along the Killing field $\xi=\frac{\partial}{\partial t}$. We note that both notions of staticity are consistent since the Ricci
1-form $Ric^{(4)}[\xi]=\xi^{\mu}Ric^{(4)}_{\,\mu\nu}\,dx^{\nu}$ is zero due to the field equations and the fact that $\xi^{\mu}\nabla^{(4)}_{\mu}\varphi={\cal L}_{\xi}\varphi= 0$.

We will adopt the following notion of asymptotic flatness. The spacetime is called asymptotically flat if there exists a compact set $K \subset M^{(3)}$ such that
$M^{(3)} - K$ is diffeomorphic to ${\mr}^{3} \backslash \bar{B}$ where $\bar{B}$ is the closed unit ball centered at the origin in ${\mr}^3$, and such that

\begin{eqnarray}
g^{(3)}= \delta + {\cal O}(r^{-1}), \;\;\; N= 1 - \frac{M}{r} + {\cal O}(r^{-2}), \;\; \varphi =\varphi_{\infty} - \frac{q}{r} + {\cal O}(r^{-2}),
\end{eqnarray}
with respect to the standard radial coordinate $r$ of $\mr^3$. Here $M$, $\varphi_{\infty}$ and $q$ are constants with $M$ and $q$ being the mass and the scalar charge.
We will consider spacetimes with $M>0$ and without loss of generality we will set $\varphi_{\infty}=0$.

The dimensionally reduced static Einstein-scalar field equations are the following

\begin{eqnarray}\label{DRFE}
&&Ric^{(3)}_{ij} = N^{-1} \nabla^{(3)}_i \nabla^{(3)}_j N + 2 \nabla^{(3)}_i\varphi \nabla^{(3)}_j\varphi ,\nonumber \\
&&\nabla^{(3)}_i \nabla^{(3)i}N=\Delta^{(3)}N=0, \\
&&\nabla^{(3)}_i\left(N \nabla^{(3)i}\varphi\right)=0 , \nonumber
\end{eqnarray}
where $\nabla^{(3)}_i$ and $Ric^{(3)}_{ij}$ are the Levi-Civita connection and the Ricci tensor with respect to the metric $g^{(3)}_{ij}$.

Now we can define the notion of photon sphere. First we give the definition of  photon surface \cite{Claudel2001},\cite{Cederbaum2014}.

\medskip
\noindent

{\bf Definition} {\it  An embedded timelike hypersurface ${\cal P}\hookrightarrow M^{(4)}$  is called a photon surface if and only if any null geodesic initially tangent to
${\cal P}$ remains tangent to ${\cal P}$ as long as it exists. }

\medskip
\noindent

The definition of  photon sphere is  a natural extension of the  definition in the pure vacuum case \cite{Cederbaum2014}.

\medskip
\noindent

{\bf Definition} {\it Let ${\cal P}\hookrightarrow M^{(4)}$ be  a photon surface. Then  ${\cal P}$  is called a photon sphere if the lapse function $N$ and the
scalar field $\varphi$  are constant along ${\cal P}$. }

\medskip
\noindent

It is worth noting that the definition of  the photon sphere is very well justified by the known exact and numerical  solutions in general relativity and  the scalar-tensor theories
as well as the studies of the gravitational lensing in these theories.

As an additional technical assumption we shall assume that the lapse function regularly foliates the region of  spacetime $M^{(3)}_{ext}$ exterior to the photon sphere.
This means that $g^{(3)}(\nabla^{(3)}N,\nabla^{(3)}N)\ne 0$  everywhere on $M^{(3)}_{ext}$. In what follows we will use the function $\rho : M^{(3)}_{ext} \to \mr^{+} $
defined by

\begin{eqnarray}
\rho=\left[g^{(3)}(\nabla^{(3)}N,\nabla^{(3)}N)\right]^{-1/2}.
\end{eqnarray}

Now let us consider the 2-dimensional intersection $\Sigma$ of the photon sphere ${\cal P}$ and the time slice $M^{(3)}$. By the definition of the photon sphere,   $\Sigma$ which is the inner boundary of $M^{(3)}_{ext}$, is given by $N=N_{0}$ for some $N_{0}\in \mr^{+}$. The metric induced on $\Sigma$ will be denoted by $\sigma$. It is not difficult to see  that our assumptions
restrict our considerations to the case of a connected photon sphere. Moreover,  all the level sets $N=const$, including $\Sigma$, are topological spheres which is a direct consequence from our assumptions.

By the maximum principle for harmonic functions and by the asymptotic behavior of $N$ for $r\to\infty$ we obtain that the values of $N$ on $M_{ext}^{(3)}$
satisfy

\begin{eqnarray}
N_{0}\le N<1.
\end{eqnarray}

\section{Functional dependence between the lapse function and the scalar field }

Here we show that there is a functional dependence between the lapse function $N$ and the scalar field $\varphi$. More precisely we have the following

\medskip
\noindent

{\bf Lemma}. {\it The lapse function and the scalar field are subject to the relation}

\begin{eqnarray}\label{Nvarphi}
\varphi= \frac{q}{M} \ln(N),
\end{eqnarray}
 {\it with $M$ and $q$ being the mass and the scalar charge respectively}.

\medskip
\noindent

{\bf Proof}. Let us first use the fact that $N$ is harmonic on $M^{(3)}_{ext}$.  Integrating $\Delta^{(3)}N=0$ on $M^{(3)}_{ext}$  and applying the Gauss theorem  we have

\begin{eqnarray}
0= \int_{M^{(3)}_{ext}} \Delta^{(3)}N \sqrt{g^{(3)}}d^3x = \oint_{S^{2}_{\infty}} \nabla^{(3)}_i N d^2\Sigma^{i} - \oint_{\Sigma} \nabla^{(3)}_i N d^2\Sigma^{i},
\end{eqnarray}
where $S^{2}_{\infty}$ is the 2-dimensional sphere at infinity. This result and the asymptotic behavior of $N$ give

\begin{eqnarray}\label{Mass}
M=\frac{1}{4\pi}\oint_{S^{2}_{\infty}} \nabla^{(3)}_i N d^2\Sigma^{i} =\frac{1}{4\pi} \oint_{\Sigma} \nabla^{(3)}_i N d^2\Sigma^{i}.
\end{eqnarray}

In the same way making use of the field equation for $\varphi$, namely $\nabla^{(3)}_{i}\left(N\nabla^{(3)i}\varphi\right)=0$, and the asymptotic behavior of $N$ and $\varphi$,   we find

\begin{eqnarray}\label{Scharge}
q= \frac{1}{4\pi} \oint_{S^{2}_{\infty}} \nabla^{(3)}_i \varphi \, d^2\Sigma^{i}= \frac{1}{4\pi} N_{0} \oint_{\Sigma} \nabla^{(3)}_i \varphi d^2\Sigma^{i}.
\end{eqnarray}

The next step is to consider $J_{i}=\varphi  \nabla^{(3)}_i N  - N\ln(N)  \nabla^{(3)}_i \varphi$. As a consequence of the field equations for $N$ and $\varphi$, it is not difficult
one to show that $ \nabla^{(3)}_i J^{i}=0$. Integrating $ \nabla^{(3)}_i J^{i}=0$ on $M^{(3)}_{ext}$, applying the Gauss theorem and taking into account the asymptotic behavior of $N$ and
$\varphi$, we obtain

\begin{eqnarray}
\oint_{\Sigma} \varphi\nabla^{(3)}_i N d^2\Sigma^{i}= \oint_{\Sigma} N\ln(N)\nabla^{(3)}_i \varphi d^2\Sigma^{i},
\end{eqnarray}
which in view of  (\ref{Scharge})  gives

\begin{eqnarray}\label{PhiPhoton}
\varphi_{0}= \frac{q}{M}\ln(N_{0}).
\end{eqnarray}

The final step is to consider the divergence identity

\begin{eqnarray}
N^{-1}\left(\omega_{i} \,\omega^{i}\right)= \nabla^{(3)}_i \left[\left(\varphi - \frac{q}{M}\ln(N)\right)\omega^{i}\right],
\end{eqnarray}
where

\begin{eqnarray}
\omega_{i}=N\nabla^{(3)}_i\varphi - \frac{q}{M} \nabla^{(3)}_i N.
\end{eqnarray}
As one can verify this identity is a consequence of the field equations for the lapse function and the scalar field. By applying the Gauss theorem to the above identity we obtain

\begin{eqnarray}
\int_{M^{(3)}_{ext}} N^{-1}\left(\omega_{i} \,\omega^{i}\right) \sqrt{g^{(3)}}d^3x =  \oint_{S^{2}_{\infty}} \left(\varphi - \frac{q}{M}\ln(N)\right)\omega_{i} \, d^2\Sigma^{i} -
\oint_{\Sigma} \left(\varphi - \frac{q}{M}\ln(N)\right)\omega_{i} \, d^2\Sigma^{i}=0 ,
\end{eqnarray}
where we have taken into account the asymptotic behavior of $N$ and $\varphi$ as well as (\ref{PhiPhoton}) in the evaluation of the surface integrals. Since $N>0$ on $M^{(3)}_{ext}$
we conclude that $\omega_{i}=N\nabla^{(3)}_i\varphi - \frac{q}{M} \nabla^{(3)}_i N=0$ on $M^{(3)}_{ext}$. Therefore we obtain $\varphi= \frac{q}{M}\ln(N) + C$ with $C$  being a constant.
From the asymptotic behavior of $N$ and $\varphi$ or from eq.(\ref{PhiPhoton}) we find that $C=0$ which proves (\ref{Nvarphi}).

\section{Some relations for the photon sphere  and inequality for the scalar charge to mass ratio}

In this section we shall derive some key relations for  the photon sphere and an important inequality for the scalar charge to mass ratio.
All these results will be directly or indirectly   used in the proof of  the main theorem. We first recall the important result of Claudel-Virbhadra-Ellis \cite{Claudel2001}:

\medskip
\noindent

{\bf Theorem} {\it Let ${\cal P}\hookrightarrow M^{(4)}$ be an embedded timelike hypersurface. Then ${\cal P}$ is a photon surface if and only if it is  totally umbilic
(iff its second fundamental form is pure trace). }

\medskip
\noindent

Denoting the metric induced on  ${\cal P}$ by $p$ the above result can be written in the form

\begin{eqnarray}\label{PEK}
{\cal K}^{\cal P}= \frac{H^{\cal P}}{3}p ,
\end{eqnarray}
where ${\cal K}^{\cal P}$ is the second fundamental form and $H^{\cal P}$ is the mean curvature of ${\cal P}$. It is easy to show that $H^{\cal P}$
is constant on  ${\cal P}$. Using the contracted Codazzi equation for $({\cal P}, p)\hookrightarrow (M^{(4)}, g^{(4)})$ we have

\begin{eqnarray}
Ric^{(4)}(X, n)= \frac{2}{3}X(H^{\cal P}),
\end{eqnarray}
where $n$ is the unit normal to ${\cal P}$ and $X$ is a vector field tangent to ${\cal P}$, i.e. $X\in \Gamma (T{\cal P})$. Taking into account the field equations
we find $Ric^{(4)}(X, n)=2X(\varphi)n(\varphi)=0$ since $\varphi$ is constant on ${\cal P}$. Hence we conclude that $X(H^{\cal P})=0$ which means that
$H^{\cal P}$ is constant on ${\cal P}$.

The same can be proven for the Ricci scalar curvature $R^{\cal P}$ of  ${\cal P}$. Applying the  Gauss equation for $({\cal P}, p)\hookrightarrow (M^{(4)}, g^{(4)})$

\begin{eqnarray}
R^{(4)} - 2 Ric^{(4)}(n,n)= R^{\cal P} - \left(Tr {\cal K}^{\cal P}\right)^2 + Tr\left( {{\cal K}^{\cal P}}\right)^2 ,
\end{eqnarray}
and making use of the field equations and  (\ref{PEK}), we find

\begin{eqnarray}
R^{\cal P}= \frac{2}{3} (H^{\cal P})^2 - 2\left(n(\varphi)\right)^2 .
\end{eqnarray}
Since $H^{\cal P}$ is constant on ${\cal P}$ we have to show that $n(\varphi)$ is  constant  on ${\cal P}$. In view of the relation (\ref{Nvarphi}), it is enough to show that
$n(N)$ is constant on ${\cal P}$. Since the spacetime is static, it is sufficient to prove that $n(N)$ is constant on $\Sigma$. This will be done below.

For the second fundamental form ${\cal K}^{\Sigma}$ of  $(\Sigma,\sigma)\hookrightarrow (M^{(3)},g^{(3)})$ (with a unit normal $n$) we have
\begin{eqnarray}
{\cal K}^{\Sigma}(X, Y)= g^{(3)}(\nabla^{(3)}_{X}n, Y)= g^{(4)}(\nabla_{X}n, Y)= {\cal K}^{\cal P}(X,Y)= \frac{H^{\cal P}}{3} p(X,Y)= \frac{H^{\cal P}}{3} p(X,Y) ,
\end{eqnarray}
where $X,Y \in \Gamma(T\Sigma)$. Therefore we find ${\cal K}^{\Sigma}=\frac{H^{\cal P}}{3}\sigma$ which also gives a simple relation between the mean curvatures $H^{\cal P}$
and $H^{\Sigma}$, namely $H^{\Sigma}=\frac{2}{3}H^{\cal P}$. Using this information in the contracted Codazzi equation we easily find that $Ric^{(3)}(X,n)=0$.

In order to show that $n(N)$ is constant on $\Sigma$ we follow \cite{Cederbaum2014}. For  an arbitrary $X\in\Gamma(T\Sigma)$ we have

\begin{eqnarray}
X(n(N))= X(n(N)) - (\nabla^{(3)}_{X}n)(N)= (\nabla^{(3)}\nabla^{(3)}N)(X,n)= \nonumber \\ N \left[Ric^{(3)}(X,n) - 2X(\varphi)n(\varphi)\right]=0,
\end{eqnarray}
where we have taken into account the dimensionally reduced field equations (\ref{DRFE}) and the fact that $N$ and $\varphi$ are constant on $\Sigma$.
Therefore $n(N)$  is indeed constant on $\Sigma$.

From the Gauss equation for $(\Sigma, \sigma)\hookrightarrow ({\cal P},p)$, and  taking into account that the spacetime is static, it is  easy to show
that the Ricci scalar curvature $R^{\Sigma}$ of $\Sigma$ is given by

\begin{eqnarray}\label{RSCS1}
R^{\Sigma}= R^{\cal P}= \frac{2}{3} (H^{\cal P})^2 - 2\left(n(\varphi)\right)^2 .
\end{eqnarray}

The Gauss equation

\begin{eqnarray}
R^{(3)} - 2 Ric^{(3)}(n,n)= R^{\Sigma} - \left(Tr {\cal K}^{\Sigma}\right)^2 + Tr\left( {\cal K}^{\Sigma}\right)^2
\end{eqnarray}
for  $(\Sigma,\sigma)\hookrightarrow (M^{(3)},g^{(3)})$ gives

\begin{eqnarray}\label{Ric1}
R^{(3)} - 2 Ric^{(3)}(n,n)=R^{\Sigma} - \frac{1}{2} (H^{\Sigma})^2 .
\end{eqnarray}

In order to find $Ric^{(3)}(n,n)$ we can use the dimensionally reduced field equations (\ref{DRFE}) and

\begin{eqnarray}
\Delta^{(3)}N= \Delta^{(2)}N + \nabla^{(3)}\nabla^{(3)}N(n,n) + H^{\Sigma}n(N),
\end{eqnarray}
which leads to

\begin{eqnarray}\label{Ric2}
N_{0} Ric^{(3)}(n,n)= -H^{\Sigma} n(N) + 2N_{0}(n(\varphi))^2.
\end{eqnarray}

Using again the dimensionally reduced field equations one can easily show that $R^{(3)}=2(n(\varphi))^2$
which combined with (\ref{Ric1}) and (\ref{Ric2}) gives

\begin{eqnarray}\label{RSCS2}
N_{0}R^{\Sigma}= 2H^{\Sigma}n(N) + \frac{1}{2} N_{0} (H^{\Sigma})^2 - 2N_{0}(n(\varphi))^2.
\end{eqnarray}

We can eliminate $R^{\Sigma}$ from  (\ref{RSCS1}) and (\ref{RSCS2}) by integrating over $\Sigma$ and using
the Gauss-Bonnet theorem $\int_{\Sigma} R^{\Sigma}\sqrt{\sigma}d^2x=8\pi$ for the topological sphere $\Sigma$.
The integration reduces  (\ref{RSCS1}) and (\ref{RSCS2}) to the following relations

\begin{eqnarray}\label{IDFHN}
&& 1= \frac{3}{16\pi}(H^{\Sigma})^2 {\cal A}_{\Sigma} - \frac{1}{4\pi} (n(\varphi))^2 {\cal A}_{\Sigma}, \\
&& N_{0}= \frac{1}{4\pi} H^{\Sigma} n(N) {\cal A}_{\Sigma} +  \frac{1}{16\pi}N_{0} (H^{\Sigma})^2 {\cal A}_{\Sigma}  -  \frac{1}{4\pi} N_{0}(n(\varphi))^2 {\cal A}_{\Sigma}.\nonumber
\end{eqnarray}

From these equations it is easy to show that the following important relation is satisfied

\begin{eqnarray}
2n(N)=N_{0}H^{\Sigma},
\end{eqnarray}
or equivalently
\begin{eqnarray}\label{HNRHO}
N_{0}H^{\Sigma}\rho_{0}=2,
\end{eqnarray}
by taking into account that $n(N)=\rho^{-1}_{0}$.

The presence of a photon sphere imposes a very important  restriction on the scalar charge to mass ratio. It is convenient to express this restriction in terms of the parameter
$\nu$ defined by

\begin{eqnarray}
\nu= \left(1 + \frac{q^2}{M^2}\right)^{-1/2}.
\end{eqnarray}
For this purpose  we first rewrite (\ref{Mass}) and (\ref{Scharge}) in the form

\begin{eqnarray}\label{Mass1}
M=\frac{1}{4\pi}n(N){\cal A}_{\Sigma} =\frac{{\cal A}_{\Sigma}}{4\pi\rho_{0}},
\end{eqnarray}

\begin{eqnarray}\label{Scharge1}
q=\frac{1}{4\pi}N_{0} n(\varphi){\cal A}_{\Sigma} .
\end{eqnarray}

Substituting  relations (\ref{Mass1}) and (\ref{Scharge1})  in (\ref{IDFHN})  and after some algebra we
obtain

\begin{eqnarray}
\frac{4\nu^2 -1}{\nu^2} = N^2_{0} \,\frac{\rho_{0}}{M}>0.
\end{eqnarray}
This inequality  combined  with the definition of $\nu$ gives the desired inequality for $\nu$, namely

\begin{eqnarray}
\frac{1}{2} < \nu \le 1,
\end{eqnarray}
which is equivalent to $\frac{q^2}{M^2}< 3$.

\section{Uniqueness theorem}

The main result of the present paper is the following

\medskip
\noindent

{\bf Theorem} {\it There can be only one static and asymptotically flat spacetime $(M^{(4)}, g^{(4)},\varphi)$,  satisfying the static Einstein-scalar field equations,
possessing a photon sphere ${\cal P}\hookrightarrow M^{(4)}$ as an inner boundary of $M^{(4)}$, with lapse function $N$ regularly foliating  $M^{(4)}$ and given mass $M$
and scalar charge $q$. Moreover, the  solution is isometric to the Janis-Newman-Winicour solution with $\frac{1}{2} < \nu(M,q)\le 1$. }

\medskip
\noindent

{\bf Proof:} The strategy of the proof is the following. The first and the most difficult step is to prove that the spacetime is spherically symmetric. Then
we construct the solution explicitly.

Let us consider the 3-metric $h$ on $M^{(3)}_{ext}$ defined by

\begin{eqnarray}\label{hmetric}
h_{ij}=N^{2}g^{(3)}_{ij}.
\end{eqnarray}

In terms of the new metric the dimensionally reduced  equations become

\begin{eqnarray}\label{DRFEH}
&&R(h)_{ij}= 2D_{i}\ln(N)D_{j}\ln(N) + 2 D_{i}\varphi D_{j}\varphi ,\nonumber\\
&&D_{i}D^{i}\ln(N) = 0, \\
&&D_{i}D^{i}\varphi = 0, \nonumber
\end{eqnarray}
where $D_{i}$ and $R(h)_{ij}$ are the Levi-Chivita connection and the Ricci tensor with respect to $h_{ij}$, respectively.  Taking into account the functional dependance
$\varphi=\frac{q}{M}\ln(N)$ the  equations (\ref{DRFEH}) can be cast in the form

\begin{eqnarray}\label{EFFE}
&&R(h)_{ij}= 2D_{i}\ln({\tilde N})D_{j}\ln({\tilde N}), \\
&&D_{i}D^{i}\ln(\tilde N) = 0, \nonumber
\end{eqnarray}
where we have introduced a new function

\begin{eqnarray}\label{newlapse}
{\tilde N}= N^{1/\nu}.
\end{eqnarray}

Proceeding further we consider the Bach tensor

\begin{eqnarray}
R(h)_{ijk}= 2D_{[i}R(h)_{j]k} + \frac{1}{2}h_{k[i}D_{j]}R(h).
\end{eqnarray}
Using eqs. (\ref{EFFE}) after a long calculation we find

\begin{eqnarray}\label{DIV1}
D^{i} \left(\Omega^{-1} D_{i}\chi \right)=\frac{1}{16}\chi^{-7}\Omega^3 R(h)_{ijk} R(h)^{ijk} ,
\end{eqnarray}
where
\begin{eqnarray}
   \chi= \left(h^{ij}D_{i}HD_{j}H \right)^{\frac{1}{4}}, \;\;\; \;  \;  H=\frac{1- {\tilde N}}{1+ {\tilde N}}, \;\; \;\; \Omega= \frac{4 {\tilde N}}{\left(1+ {\tilde N}\right)^2}.
\end{eqnarray}

We can combine the equation for ${\tilde N}$ from (\ref{EFFE}) with eq. (\ref{DIV1}) to obtain another divergence identity

\begin{eqnarray}\label{DIV2}
D^{i} \left[\Omega^{-1}\left(HD_{i}\chi - \chi D_{i}H\right) \right]= \frac{1}{16} H \chi^{-7}\Omega^3 R(h)_{ijk} R(h)^{ijk}.
\end{eqnarray}

Since $H>0$ we obtain  the inequality

\begin{eqnarray}\label{II1}
\int_{M^{(3)}_{ext}} D^{i} \left[\Omega^{-1}\left(HD_{i}\chi - \chi D_{i}H\right) \right] \sqrt{h} d^3x \ge 0.
\end{eqnarray}

Another inequality can be obtained by taking into account that $H<1$, namely

\begin{eqnarray}\label{II2}
\int_{M^{(3)}_{ext}} D^{i} \left(\Omega^{-1} D_{i}\chi \right) \sqrt{h} d^3x \ge \int_{M^{(3)}_{ext}} D^{i} \left[\Omega^{-1}\left(HD_{i}\chi - \chi D_{i}H\right) \right] \sqrt{h} d^3x .
\end{eqnarray}

In both cases (\ref{II1}) and (\ref{II2}) the equality holds if and only if $R(h)_{ijk}=0$.

Calculating the integral in  (\ref{II1}) by using the Gauss theorem and taking into account (\ref{HNRHO}) we get the inequality

\begin{eqnarray}
{\tilde N}^2=N_{0}^{\frac{2}{\nu}}\le \frac{2\nu-1}{2\nu+ 1}.
\end{eqnarray}
In the same way (\ref{II2}) gives

\begin{eqnarray}
{\tilde N}^2=N_{0}^{\frac{2}{\nu}}\ge \frac{2\nu-1}{2\nu+ 1}.
\end{eqnarray}
Hence we conclude that ${\tilde N}^2=N_{0}^{\frac{2}{\nu}}= \frac{2\nu-1}{2\nu+ 1}$ and therefore $R(h)_{ijk}=0$. This means that the metric $h_{ij}$ is conformally flat.
As an immediate consequence we get that $g_{ij}$ is conformally flat too (i.e. $R(g)_{ijk}=R^{(3)}_{ijk}=0$).

Since $N$ regularly foliates $M^{(3)}_{ext}$ we can introduce adapted coordinates in which the metric $g^{(3)}_{ij}$ takes the form

\begin{eqnarray}
g^{(3)}= \rho^2 dN^2 + \sigma ,
\end{eqnarray}
where $\sigma_{AB}$ is the 2-dimensional metric on the 2-dimensional intersections ${\Sigma_{N}}$ of the level sets $N=const$  with the time slice $M^{(3)}_{ext}$. Using the formula

\begin{eqnarray}
R^{(3)}_{ijk}  R^{(3)ijk}= \frac{4}{N^4\rho^4}\left[\left({\cal K}^{\Sigma_{N}}_{AB}- \frac{1}{2}H^{\Sigma_{N}} \sigma_{AB}\right)
\left({\cal K}^{\Sigma_{N}\, AB}- \frac{1}{2}H^{\Sigma_{N}} \sigma^{AB} \right) + \frac{1}{\rho^2}\sigma^{AB}\partial_{A}\rho\partial_{B}\rho\right],
\end{eqnarray}
 where ${\cal K}^{\Sigma_{N}}_{AB}$ is the extrinsic curvature of $(\Sigma_{N},\sigma)\hookrightarrow (M^{(3)},g^{(3)})$  and $H^{\Sigma_{N}}$ is its trace, we
 conclude that

\begin{eqnarray}
{\cal K}^{\Sigma_{N}}_{AB}= \frac{1}{2}H^{\Sigma_{N}} \sigma_{AB},\;\;\; \partial_{A}\rho=0.
\end{eqnarray}
The last equality means that $\rho$ depends only on $N$. As in the previous section one can show that $H^{\Sigma_{N}}$ and the Ricci scalar curvature  $R^{\Sigma_{N}}$ are  constant on $\Sigma_{N}$. Therefore $\Sigma_{N}$ are round spheres. Hence it is easy to conclude that the metric $g^{(3)}_{ij}$ is spherically symmetric. The same holds for the metric
$h_{ij}$.

In order to find the solution in explicit form let us notice that equations  (\ref{EFFE}) are in fact the static vacuum Einstein equations written in terms of the metric $h_{ij}$
with an effective lapse function ${\tilde N}$ having the asymptotic ${\tilde N}= 1 - \frac{M}{\nu r} + O(r^{-2})$. Knowing that the Schwarzschild solution is the only static and spherically symmetric solution of the vacuum Einstein equations we have

\begin{eqnarray}
{\tilde N}^2=1 - \frac{2M}{\nu r}, \;\;\;\; h_{ij}dx^{i}dx^{j}= \left(1 - \frac{2M}{\nu r}\right)\left[\frac{dr^2}{1- \frac{2M}{\nu r}} + r^2 d\Omega^2_{S^2}\right] ,
\end{eqnarray}
where $d\Omega^2_{S^2}$ is the standard metric on the unit 2-dimensional sphere. Recovering the original lapse function $N$  from (\ref{newlapse}) and the 3-metric form (\ref{hmetric})
we  find

\begin{eqnarray}
&&N^2={\tilde N}^{2\nu}= \left(1 - \frac{2M}{\nu r}\right)^{\nu}, \\
&&g^{(3)}_{ij}dx^idx^j= N^2 h_{ij}dx^i dx^j=  \left(1 - \frac{2M}{\nu r}\right)^{-\nu} dr^2  +  \left(1 - \frac{2M}{\nu r}\right)^{1-\nu} r^2 d\Omega^2_{S^2}.
\end{eqnarray}

Having the explicit expressions for the lapse function and the 3-metric we can write the desired solution

\begin{eqnarray}
&&ds^2_{JNW}=g^{(4)}_{\mu\nu}dx^{\mu}dx^{\nu}= - \left(1 - \frac{2M}{\nu r}\right)^{\nu} dt^2 + \left(1 - \frac{2M}{\nu r}\right)^{-\nu} dr^2  +  \left(1 - \frac{2M}{\nu r}\right)^{1-\nu} r^2 d\Omega^2_{S^2}, \nonumber\\
&&\varphi_{JNW}= \frac{q\nu}{2M}\ln\left(1 - \frac{2M}{\nu r}\right),
\end{eqnarray}
which is just the Janis-Newman-Winicour solution with $\frac{1}{2} < \nu(M,q)\le 1$. This completes the proof of the theorem.

In the context of scalar-tensor theories of gravity the  above  uniqueness theorem can be formulated as follows. Let us consider a scalar-tensor theory
defined by the coupling function $A(\varphi)$. Then the unique solution  possessing a photon sphere and with mass $M$ and scalar-charge $q$
is given in the Jordan frame by the metric

\begin{eqnarray}
ds^2= A^2(\varphi_{JNW})ds^2_{JNW}.
\end{eqnarray}

\section{Discussion}

In the present paper we have proved that static and asymptotically flat solutions to the Einstein-scalar field equations possessing a photon sphere are uniquely specified by their mass and scalar charge being isometric to the Janis-Newman-Winicour solution with the same mass and scalar charge subject to the constraint $\frac{1}{2} < \nu(M,q)\le 1$. Our result was proven under the technical assumption  that the lapse function regularly foliates the spacetime. This condition is rather natural from a physical point of view. Nevertheless, in seeking mathematical generality,  the mentioned condition can be relaxed. Once having the explicit form of the unique 
solution, this could be done by using a technically modified version of the approach of \cite{Bunting1987} or \cite{Miao2005}. 

In general,  in order to apply the approach of \cite{Bunting1987} or \cite{Miao2005} we have to know the unique solution in advance. In the simplest cases the unique solution can be guessed but in the more complex situations this is practically impossible. The wise approach to the problem seems to be the following. We can first use a constructive method, like the one used in the present paper, which allows us not only to prove the uniqueness  but also to derive the desired solution and then, in order to relax a given condition, we could use a technically modified  version of the approach in \cite{Bunting1987} or \cite{Miao2005}.  In  our next works we shall demonstrate this combined approach on the  Einstein-Maxwell equations and on other equations of physical interest.

\medskip
\noindent

\noindent {\bf Acknowledgements:} The author would like to thank C. Cederbaum for  discussions
and  the Research Group Linkage Programme of the Alexander von Humboldt Foundation
for the support. The networking support by the COST Action MP1304 is also gratefully acknowledged.

\end{document}